\documentclass[prd,aps,preprint,nofootinbib,superscriptaddress]{revtex4}
\usepackage{amsmath}
\usepackage{epsfig}

\begin{document}

%\hfill{NPAC-08-24}

\title{The impact of a strongly first-order phase transition\\ on the abundance of thermal relics}

\author{Carroll Wainwright}
\email{cwainwri@ucsc.edu} \affiliation{Department of Physics, University of
California, 1156 High St., Santa Cruz, CA 95064}
\author{Stefano Profumo}
\email{profumo@scipp.ucsc.edu} \affiliation{Department of Physics, University of
California, 1156 High St., Santa Cruz, CA 95064} \affiliation{Santa Cruz Institute for Particle Physics, Santa Cruz, CA 95064}

\date{\today}

\begin{abstract}

\noindent We study the impact of a strongly first-order electro-weak phase transition on the thermal relic abundance of particle species that could constitute the dark matter and that decoupled before the phase transition occurred. We define a dilution factor induced by generic first-order phase transitions, and we explore the parameter space of the minimal supersymmetric extension to the Standard Model to determine which phase transition temperatures and dilution factors are relevant for the lightest neutralino as a dark matter candidate. We then focus on a specific toy-model setup that could give rise to a strongly first-order electro-weak phase transition, and proceed to a detailed calculation of dilution factors and transition temperatures, comparing our findings to actual neutralino dark matter models. Typical models that would produce an excessive thermal relic density and that can be salvaged postulating a strongly first-order electro-weak phase transition include massive (multi-TeV) wino or higgsino-like neutralinos, as well as bino-like neutralinos in a wider mass range, with masses as low as 400 GeV. If LHC data indicate an inferred thermal neutralino relic abundance larger than the cold dark matter density, the mismatch could thus potentially be explained by electro-weak scale physics that will also be thoroughly explored with collider experiments in the near future.

\end{abstract}

\maketitle

%%%%%%%%%%%%%%%%%%%%%%%%%%%%%%%%%%%%%%%%%%%%%%%%%%%%%%%%%%%%%%%%%%%%
\section{Introduction}
%{This section by SP} 

The particle content of the Minimal Supersymmetric extension of the Standard Model (MSSM) of particle physics (see e.g. \cite{2006wss..book.....B}) provides a stable weakly interacting massive particle which is, in principle, a viable particle dark matter candidate: the lightest neutralino. If the latter is the lightest $R$-parity odd particle, its thermal relic abundance can be close to the inferred density of dark matter on cosmological scales \cite{Komatsu:2008hk}, whose ratio to the universal critical density is estimated to be $\Omega_{\rm astro}=\rho_{\rm DM}/\rho_c=0.113\ h^{-2}$ \cite{Dunkley:2008ie}, with $h$ the present day Hubble expansion rate normalized in units of 100 km per sec per Mpc.

Several studies have pointed out that the thermal relic abundance of neutralinos in MSSM models where they are the lightest supersymmetric particles actually ranges over several orders of magnitude, simple estimates of their relic abundance being therefore only a generic order-of-magnitude estimate (see e.g. \cite{Profumo:2004at, Gelmini:2006pq}). Specifically, if the lightest neutralino has unsuppressed couplings to gauge bosons, and is heavy enough for annihilation to massive weak interaction gauge bosons to be kinematically open, it rapidly annihilates into $W^+W^-$ and/or $ZZ$ final states, and its relic density $\Omega_{\rm particle}$ is {\em below} the dark matter density, at least for neutralinos lighter than about a TeV. If the lightest neutralino has suppressed couplings to gauge bosons, as is the case for bino-like or singlino-like neutralinos, or if neutralinos are very heavy, the neutralino relic density is instead typically much larger than the universal dark matter density.

MSSM models with under-abundant neutralino relic density are phenomenologically perfectly viable: the lightest neutralino can very well not be the only contributor (or it can be a sub-dominant contributor) to the universal dark matter, in a standard cosmological setup. While modified cosmological setups can be concocted to enhance the thermal relic abundance of neutralinos in those cases (for instance in the presence of non-thermal production \cite{Gelmini:2006pq}, or of a faster expansion rate at the epoch of the lightest neutralino freeze-out \cite{Salati:2002md, Profumo:2003hq}), MSSM models with {\em over-abundant} relic neutralinos are, in principle, ruled out. A caveat to this conclusion is the possibility of an episode of entropy injection occurring at temperatures below the decoupling of neutralinos from the universe's thermal bath (entropy injection before freeze-out would not change today's relic abundance of neutralinos). 

Ref.~\cite{Gelmini:2006pq} gives a thorough discussion of such a class of scenarios for the specific case of an additional scalar field that drives both the entropy dilution and possibly the non-thermal production of neutralinos, as well as a modified cosmological expansion rate if the energy density associated to the scalar field dominates the universal energy budget. Specific examples of setups that can produce a dilution in the relic abundance of neutralinos (or, for that matter, of any other thermal relics decoupled from the thermal bath before entropy injection) include for example models with moduli \cite{Moroi:1999zb} or Q-ball decays \cite{Fujii:2002kr}, and scenarios with low-temperature or even weak-scale inflation \cite{Lyth:1995ka, Knox:1992iy, Nardini:2007me}. While a phase transition that plausibly occurred {\em after} the neutralino freeze-out is the QCD phase transition, lattice simulations \cite{Fukugita:1991mk} clearly indicate that the transition is not strongly first-order (see e.g. the discussion in Ref.~\cite{Jungman:1995df}).

In the present study, we consider a specific instance of a possible source of dilution for the thermal relic abundance of relic neutralinos from the early universe, namely the electro-weak phase transition (EWPT). In the minimal formulation of the Standard Model Higgs sector, the EWPT is very weakly first-order, or entirely absent, as found e.g. in the non-perturbative analysis of Ref.~\cite{Rummukainen:1998as}. However, a strongly first-order phase transition can occur within the MSSM, or even in very simple extensions of the scalar sector of the Standard Model (for instance via the inclusion of additional singlet scalars \cite{singlet}). Other possibilities include for instance models with multiple hidden sector scalars coupled only to the Higgs sector (which thus acts as a ``portal'' to the hidden sector, as envisioned in Ref.~\cite{Schabinger:2005ei, Patt:2006fw, Bowen:2007ia}). This possibility, which includes scenarios with tree-level conformal invariance where the Higgs mass is generated via dimensional transmutation, has been discussed and studied in detail in Ref.~\cite{Espinosa:2007qk} and \cite{Espinosa:2008kw}. A strongly first-order phase transition needs to actually be posited in the context of scenarios where the baryon asymmetry is produced at the EWPT, to prevent the ``washout'' of the generated baryon number density by sphaleron processes (for a pedagogical introduction to the framework of electro-weak baryogenesis see e.g. Ref.~\cite{Quiros:1999jp}; classic studies on MSSM baryogenesis include \cite{Carena:2002ss} and \cite{Konstandin:2005cd}; recent discussions of the phenomenology of electro-weak baryogenesis in the MSSM are given in Ref.~\cite{Cirigliano:2006dg, Carena:2008vj}).  In this case, the order parameter can be considered to be the ratio of the SU(2) Higgs background field $\phi=\sqrt{2}\langle H^0\rangle$ to the critical temperature for the phase transition, $T_c$, and a strongly first-order phase transition corresponds to $\phi_c/T_c\gtrsim1$ \cite{Quiros:1999jp}.

The entropy injection produced in any first-order phase transition can play a relevant role in the thermal history of species that froze out prior to the phase transition, since their relic density will be diluted away by an amount dependent upon the relative entropy injected to the entropy in the species in thermal equilibrium. Interestingly, the CERN Large Hadron Collider (LHC) might potentially inform us on both the details of the scalar electro-weak sector and of the EWPT in particular (including possible non-minimal extensions of the scalar sector \cite{singlet}), as well as on the mass of a putative dark matter particle. In some cases, data from the LHC might even enable us to infer, via the knowledge of particle masses and couplings, the thermal relic abundance of a stable neutralino that might be the main dark matter constituent (explicit examples with realistic assumptions on the LHC performance are given in Ref.~\cite{Baltz:2006fm}). A mismatch between the observed cold dark matter abundance and the inferred relic abundance of candidate particles discovered with colliders can have very profound consequences. If the resulting inferred thermal abundance were {\em larger} than the cold dark matter density, then the answer to the ensuing conundrum might lie in the EWPT and in the Higgs sector, and thus again in physics that can, and will, be tested with the LHC.

The present study is organized as follows: in the next section we describe in detail how a first-order phase transition affects the abundance of a species that froze out prior to the temperature at which the phase transition occurred; the following sec.~\ref{sec:mssm} explores the ranges of freeze-out temperatures and of relic abundances relevant to the case of the lightest neutralinos of the MSSM. The ensuing sec.~\ref{sec:dilution} describes simple models for the EWPT, and sec.~\ref{sec:EWPT-SUSY} describes their impact upon the relic density of MSSM neutralinos.  These last two sections present our main results. Finally, sec.~\ref{sec:concl} summarizes and concludes.

%%%%%%%%%%%%%%%%%%%%%%%%%%%%%%%%%%%%%%%%%%%%%%%%%%%%%%%%%%%%%%%%%%%%
\section{Relic Density Dilution: Thermodynamics}

We are concerned here with determining the dilution of a relic species that decoupled (or ``froze-out'') from the thermal bath in the early universe by entropy injected during a first-order phase transition occurring after the species' freeze-out.  The basic thermodynamics of the dilution, which we describe in the present section, is fully general, and does not depend upon the specifics of the model for the phase transition.

Let $\mathcal{F}(\phi, T)$ be the finite temperature effective potential of the early universe, where $\phi$ is the order parameter of the phase transition, for instance the vacuum expectation value of the neutral component of the Higgs field.  We assume that, for the temperatures under consideration here, there are at most two minima of the potential.  At very high temperatures, the potential only has one minimum, at $\phi = 0$.  As the temperature drops, a second minimum develops with a corresponding effective potential value larger than that at $\phi = 0$. 
% Since at $T=0$ we want spontaneous symmetry breaking, hence the second minimum must be the absolute minimum, the value of the potential at the second minimum relative to the first will eventually decrease with decreasing temperature. 
The value of the potential at the second minimum relative to the first decreases with decreasing temperature so that at $T=0$ the second minimum is the absolute minimum and there is spontaneous symmetry breaking.
We define a ``critical temperature'' as the temperature $T = T_c$ when the two minima are degenerate.  At some temperature $T_* < T_c$, the system transitions from $\phi = 0$ to the new minimum: the tunneling probability from the $\phi=0$ minimum to the true vacuum is of order unity.

First, let us assume that the transition temperature $T_*$ is very close to the critical temperature $T_c$ so that supercooling is negligible.  Here, we follow the discussion of M\'egevand and S\'anchez \cite{Megevand:2007sv} (see also \cite{Anderson:1992}).  Let $s_+$ be the entropy density of the high-temperature phase, and let $s_-$ be the entropy density of the low-temperature phase.  We may then write the total entropy density as
\begin{equation}
\label{eq:s}
s = s_+ - f \Delta s,
\end{equation}
where $\Delta s = s_+ - s_-$ and $f$ is the volumetric fraction of the system in the low temperature phase.  Since the minima are degenerate at $T_c$, the system is in equilibrium and the total entropy is conserved.  The entropy density then scales as
\begin{equation}
\label{eq:s2}
s = s_+ \left(\frac{a_i}{a}\right)^3,
\end{equation}
where $a_i$ is the scale factor {of the universe} at the beginning of the phase transition.  The transition is complete once $f = 1$.  Plugging this into Eq. (\ref{eq:s}) and combining with Eq. (\ref{eq:s2}), one finds that the total expansion during the phase transition is
\begin{equation}
\label{eq:dilution1}
\left(\frac{a_f}{a_i}\right)^3 = \frac{1}{1 - \Delta s/s_+} = \frac{s_+}{s_-}.
\end{equation}
This is the equation we use to find the dilution in sec.~\ref{sec:semi-analytic}.  

Realistically, the system would not be in exact equilibrium---friction and collisions in the walls of bubbles of true vacuum would release entropy, slightly increasing the dilution \cite{Ignatius:1993qn}.  This can be quantitatively captured by assessing the variation of the dilution factor with a variation to the temperature at which the phase transition occurs, this variation being driven by the mentioned effects (bubble collisions, friction in the bubble walls etc.). If the entire transition occurs at a temperature an amount $\Delta T$ below $T_c$, then one can show that the change in dilution $D$ is $\Delta D = -3D(D-1)(\Delta T/T_c)$, assuming a radiation dominated energy density.
%This is also the amount of dilution experienced by relic species during the transition.

If $T_* \ll T_c$, then the two phases are not in equilibrium at the beginning of transition and Eq. (\ref{eq:s2}) does not hold.  Instead, there are three distinct stages to the phase transition: a supercooling stage, a reheating stage, and a phase-coexistence stage.  During supercooling, entropy is conserved, so that
\begin{equation}
\label{eq:s-cool}
\frac{s_+(T_*)}{s_+(T_c)} = \left(\frac{a_i}{a_*}\right)^3
\end{equation}
where $a_*$ is the scale factor at the minimum of supercooling.  Assuming that reheating happens quickly relative to the expansion rate, the energy density $\rho$ of the universe does not change during reheating.  The entropy, however, does.  If {a large enough amount of reheating occurs}, the universe will reach a phase coexistence stage at $T=T_c$.  Conservation of energy then gives the initial fraction of the universe in the low-temperature phase at the beginning of phase coexistence:
\begin{align}
\begin{split}
\rho_+(T_*) &= \rho_+(T_c) - f_0 [\rho_+(T_c) - \rho_-(T_c)] \\ 
&= \rho_+(T_c) - f_0 L 
\end{split} \\
\rightarrow f_0 &= \frac{ \rho_+(T_c) -  \rho_+(T_*)}{L},
\label{eq:f0}
\end{align}
where $L$ is the latent heat of the transition at $T_c$.  The entropy density during phase-coexistence is then
\begin{equation}
\label{eq:s2-cool}
s = (s_+ - f_0 \Delta s) \left(\frac{a_*}{a}\right)^3.
\end{equation}
Combining this with Eq.~(\ref{eq:s}), we have the expansion during phase coexistence
\begin{equation}
\label{eq:dilution2}
\left(\frac{a_f}{a_*}\right)^3 = \frac {1 - f_0 \Delta s / s_+(T_c)} {1 - \Delta s / s_+(T_c)},
\end{equation}
which gives a total expansion of 
\begin{equation}
\label{eq:dilution3}
\left(\frac{a_f}{a_i}\right)^3 = \left(\frac {1 - f_0 \Delta s / s_+(T_c)} {1 - \Delta s / s_+(T_c)}\right) \frac{s_+(T_c)}{s_+(T_*)}.
\end{equation}
We use this equation in sec.~\ref{sec:numeric-dilution} where the transition is strongly first-order.

\subsection{Determining the transition temperature}
{We specify here the definition we adopt for the transition temperature $T_*$, given its relevance in determining the dilution factor for a given effective potential}. Shortly after the universe cools below the critical temperature, bubbles of true vacuum nucleate via thermal tunneling.  Most of these are too small to grow---the pressure difference $\Delta p = -\Delta\mathcal{F}$ between the true and false vacuum is not {large} enough to overcome the surface tension of their walls, so they collapse.  Only large bubbles can grow.  As the universe {further} cools, the nucleation rate of larger bubbles increases dramatically.  The phase transition begins once the probability to nucleate a supercritical bubble in one Hubble volume is of order 1.  Tunneling in cosmological phase transition was originally discussed in the seminal work of Ref.~\cite{coleman,linde} (see also Ref.~\cite{Anderson:1992}). For a pedagogical review of cosmological phase transition, see e.g. Kolb and Turner \cite{kt}.

The tunneling probability per unit time per unit volume {goes with temperature} roughly as $\Gamma \sim T^4\exp^{-S_3/T}$, where $S_3$ is the three-dimensional Euclidean action
\begin{equation}
\label{eq:S3}
S_3 = 4\pi \int_0^\infty r^2dr\left[ \frac{1}{2}\left(\frac{d\phi}{dr}\right)^2+\mathcal{F}(\phi(r),T)\right]
\end{equation}
and {where} we assume spherical symmetry.  The bubble shape $\phi(r)$ comes from the corresponding Euclidean equation of motion
\begin{equation}
\label{eq:euclid}
\frac{d^2\phi}{dr^2} + \frac{2}{r}\frac{d\phi}{dr} = \frac{\partial}{\partial\phi}\mathcal{F}(\phi,T),
\end{equation}
with the boundary conditions $\lim_{r\to\infty}\phi(r) = 0$ and $\left.\frac{d\phi}{dr}\right|_{r=0}=0$.  Finally, the requirement that one supercritical bubble nucleates per horizon volume in a Hubble time yields a phase transition temperature $T_*$ such that 
\begin{equation}
\label{eq:T*}
\int_{T_*}^\infty \frac{dT}{T}\left(\frac{2\zeta M_{Pl}}{T}\right)^4\exp^{-S_3(T)/T} = \mathcal{O}(1),
\end{equation}
where $M_{Pl}$ is the Planck mass, $\zeta = \frac{1}{4\pi}\sqrt{\frac{45}{\pi g}}$, and $g$ is the effective number of relativistic degrees of freedom (see Ref.~\cite{Quiros:1999jp}). For temperatures {around} the electroweak scale, Eq.~(\ref{eq:T*}) implies that $S_3/T \sim \mathcal{O}$(130--140).

%%%%%%%%%%%%%%%%%%%%%%%%%%%%%%%%%%%%%%%%%%%%%%%%%%%%%%%%%%%%%%%%%%%%
\section{Salvaging MSSM models with over-abundant relics}\label{sec:mssm}

In this section we calculate the degree of relic density dilution needed in MSSM models as a function of the temperature at which neutralinos freeze out. This will allow us to immediately determine whether a given MSSM model can or cannot be salvaged by a strongly first-order electro-weak phase transition, the requirement being that the critical temperature of the phase transition be smaller than the freeze-out temperature, and the dilution factor be large enough to bring the neutralino relic density at or below the level of the inferred average dark matter density in the universe.

We work in the context of the $R$-parity, flavor and CP-conserving MSSM, we enforce that the lightest supersymmetric particle be the lightest neutralino, and parameterize the soft supersymmetry breaking parameters that enter the relevant particle spectrum for the calculation of the neutralino relic abundance with their values at the electro-weak scale (we thus do not assume any grand unified structure for the soft supersymmetry-breaking terms). We do not assume any relationship between the gaugino soft supersymmetry breaking masses, nor about their relative signs, but we assume a common mass scale for all sfermions. Also, for simplicity we set to zero all trilinear scalar coupling with the exception of third generation sfermions. The gluino mass (which does not enter into the calculation of the relic abundance in the DarkSUSY code) was set to $M_3 = 3M_2$, which approximately follows the usual supergravity relation.  Details of our scan procedure, including lower and upper limits for the scan as well as whether the sampling was carried out logarithmically or linearly, are given in Table~\ref{tab:scan}. %{\bf To avoid numerical issues in calculating the relic abundance of MSSM neutralinos, we also enforce the mass of the lightest neutralino to be below 10 TeV}.
%----------------------------------------------------------------
\begin{table}[!t]
\begin{center}
\begin{tabular}{|c|c|c|c|}
\hline
Parameter & Lower Lim. & Upper Lim. & Scan Type \\
\hline\hline
$|\mu|$ & 66 GeV & 20 TeV & Log \\
$|M_1|$ & 40 GeV & 20 TeV & Log \\
$|M_2|$ & 83 GeV & 20 TeV & Log \\
$m_{\tilde f}$ & $\tfrac{3}{2}$min($M_1, M_2, \mu$) & 20 TeV & Log \\
$A_{\tilde f}$ & -3 & 3 & Lin \\
$m_A$ & 200 GeV & 20 TeV & Log \\
$\tan\beta$ & 2.5 & 60 & Lin \\
\hline
\end{tabular}
\label{tab:scan}\caption{\it\small Ranges for the parameter space scan we employ for our fig.~\ref{fig:mssm}.}
\end{center}
\end{table}
%----------------------------------------------------------------

Once a particular MSSM setup is defined by the random parameter values picked by the procedure outlined above, we require that the resulting particle setup be compatible with updated versions of the limits from collider searches, rare decays and electro-weak precision measurements described in Ref.~\cite{Gondolo:2004sc}. We then calculate the thermal relic abundance of the lightest neutralinos, in the context of a standard cosmological setup, with the DarkSUSY package \cite{Gondolo:2004sc}.

The two parameters we are interested in for the present study are the thermal neutralino relic abundance, and the temperature at which the neutralino freezes out: a phase transition occurring at temperatures larger than the freeze-out temperature would not affect the relic abundance of neutralinos, since the latter would be in thermal equilibrium after the phase transition and its number density would re-equilibrate to the other thermal species. We define the freeze-out temperature according to the prescription of Ref.~\cite{Gondolo:1990dk}, that sets it as the temperature where the comoving neutralino number density is a factor 2.5 larger than its asymptotic zero-temperature value.

%----------------------------------------------------------------
\begin{figure}[!t]
\begin{center}
\hspace*{-1.cm}\psfig{file=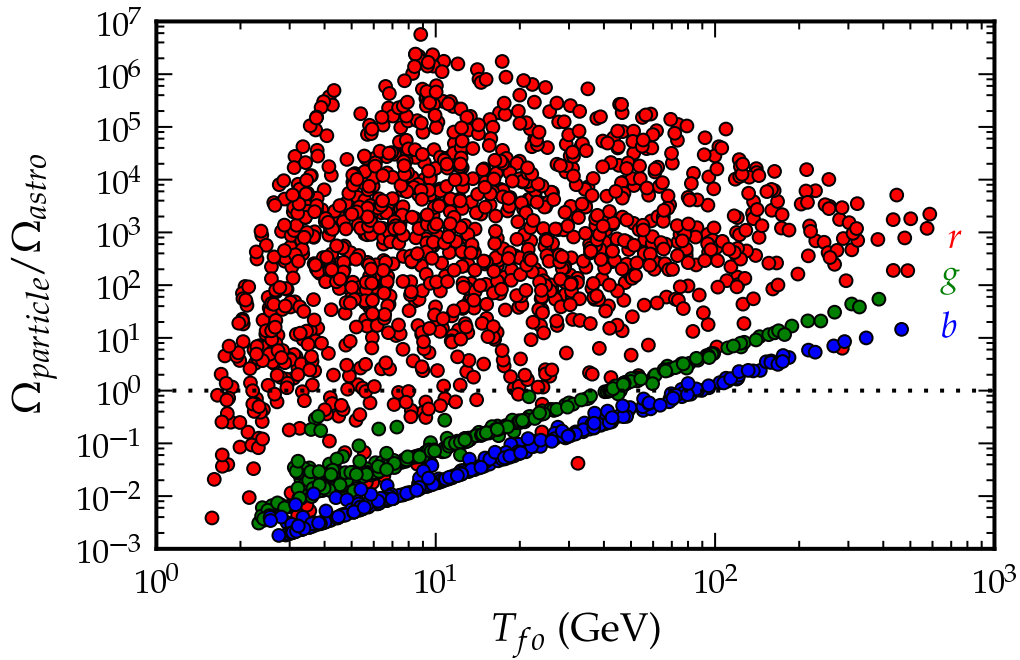,width=0.85\textwidth}
\end{center}
\caption{\it\small A scatter plot of MSSM models on the plane defined by the ratio of the models' thermal neutralino relic abundance ($\Omega_{\rm particle}$) over the universal dark matter density ($\Omega_{\rm astro}$), versus the neutralino freeze-out temperature $T_{fo}$. Red points correspond to bino-like lightest neutralino, while blue and green points to wino- and higgsino-like lightest neutralinos, respectively. Models above the horizontal line at $\Omega_{\rm particle}/\Omega_{\rm astro}=1$ are over-abundant, and are ruled out unless a dilution mechanism such as the one we discuss in the present study is operative. See the text for definitions and details.}
\label{fig:mssm}
\end{figure}
%----------------------------------------------------------------

% Using Omega_astro = .23, h (hubble) = .71
Fig.~\ref{fig:mssm} shows the minimal dilution factor required to bring the thermal neutralino relic abundance below the upper limit set by determinations of the average universal dark matter density, $\Omega_{\rm astro}h^2 \simeq 0.113$ \cite{Dunkley:2008ie} as a function of the neutralino freeze-out temperature. Loosely, the latter is a factor of 20-25 times smaller than the neutralino mass, which thus increases linearly with the x-axis. We indicate with blue dots models where $M_2<M_1,\ \mu$, i.e. models corresponding to a dominant wino component in the lightest neutralino mass eigenstate; green dots correspond to models with $\mu<M_1,\ M_2$ (higgsino-dominated lightest neutralinos) and red dots to $M_1<M_2,\ \mu$ (bino-dominated lightest neutralinos). 

Notice that the pair annihilation of bino-like neutralinos proceeds through a variety of channels, including squark and higgs exchanges, which dramatically depend on the details of the spectrum of the relevant particles. This induces a wide scatter in the relic abundance versus mass (or freeze-out temperature) of bino-like models, as evident in fig.~\ref{fig:mssm}. On the other hand, for dominantly wino- and higgsino-like neutralinos, the dominant annihilation modes always proceed through gauge boson pairs, mediated by chargino (or by the next-to-lightest higgsino in the case of $ZZ$ final state) exchange. In this case, the pair-annihilation cross section is fixed by gauge couplings and by the mass of the neutralino/chargino system, modulo kinematic threshold effects for neutralino masses near the gauge boson mass (this shows up in fig.~\ref{fig:mssm} for $T_{\rm fo}\sim2-5$ GeV). This is the reason why the wino and the higgsino-like models all fall approximately on a line in the log-log plot of fig.~\ref{fig:mssm}.

MSSM models whose relic abundance can be salvaged by a strongly first-order electro-weak phase transition therefore feature either higgsino-like lightest neutralinos with masses larger than a TeV, or wino-like neutralinos with masses in excess of 2 TeV, or bino-like neutralinos in a wide mass range, provided the freeze-out temperature is large enough to be above the temperature at which the phase transition occurs. We discuss the possible effects of entropy dilution in the electro-weak phase transition and its impact on the MSSM parameter space in the following section, {where we show which regions of fig.~\ref{fig:mssm} can potentially be salvaged by the resulting entropy injection}.

%%%%%%%%%%%%%%%%%%%%%%%%%%%%%%%%%%%%%%%%%%%%%%%%%%%%%%%%%%%%%%%%%%%%
\section{Relic Density Dilution and First-Order Phase Transitions}\label{sec:dilution}

\subsection{Overview of the Field-Theoretical Setup}

We calculate here the dilution of a thermal relic due to a first-order phase transition driven by a scalar field $\phi$.  Since we are interested in the specific case of the EWPT, as an illustrative instance we consider here the effective potential  of the neutral component of the scalar electro-weak sector of the Standard Model $\phi$. The extension to the Standard Model we then take into consideration will include an additional set of fermionic and/or bosonic degrees of freedom, which might be thought of as being singlets under the Standard Model gauge group, but that might transform nontrivially under a hidden sector gauge group, and are coupled to the visible sector only through the Higgs sector \cite{Schabinger:2005ei, Patt:2006fw, Bowen:2007ia,Espinosa:2007qk,Espinosa:2008kw}. These class of models includes so-called Higgs Portal scenarios, where (partly) secluded hidden sectors communicate with the Standard Model (the ``visible'' sector) exclusively via interactions in the Higgs sector \cite{pattwil}. Early studies on such models include those listed in Ref.~\cite{higgsportal}, where the Higgs portal was often invoked in the context of identifying a viable particle dark matter candidate. The existence of hidden sectors that might be only partially secluded is ubiquitous to extension to the Standard Model, such as those based on high-rank GUT models, e.g. where the GUT group is $E_6$ \cite{e6models}, in gauge-mediated supersymmetry breaking setups (for a review, see e.g. \cite{gaugemed}) as well as in the landscape of string compactifications \cite{strings}. In all of these setups, a potentially large number of (fermionic or bosonic) degrees of freedom, neutral under the Standard Model gauge interactions, might interact with the Higgs sector alone.

%As an illustrative case, we adopt the same setup as in M\'egevand and S\'anchez \cite{Megevand:2007sv}, which corresponds to the tree-level, zero-temperature potential of the neutral component of the scalar electro-weak sector of the Standard Model, to which an additional set of fermionic or bosonic degrees of freedom is further added.  
At tree-level, the effective potential associated to the field $\phi$ only we consider here is:
\begin{equation}
\label{eq:V0}
V_0(\phi) = -\tfrac{1}{2}\lambda v^2 \phi^2 + \tfrac{1}{4}\lambda \phi^4,
\end{equation}
while the zero-temperature one-loop corrections read:
\begin{equation}
\label{eq:V1}
V_1(\phi) = \sum_i \pm \frac{g_i}{64\pi^2}\left[ m_i^4(\phi) \left(\log\left(\frac{m_i^2(\phi)}{m_i^2(v)}\right) - \frac{3}{2}\right) + 2m_i^2(\phi)m_i^2(v)\right],
\end{equation}
where $g_i$ is the number of degrees of freedom of each particle species in the theory coupled to $\phi$, $m_i(\phi)$ is the particle species mass, and the upper and lower signs correspond to bosons and fermions, respectively. We consider only scalar bosons in this paper. Notice that the potential for vector bosons would carry a constant of $\tfrac{5}{6}$ in place of the $\tfrac{3}{2}$ in Eq.~(\ref{eq:V1}) (see Ref.~\cite{Espinosa:2007qk}).
Note that both $V_0(\phi)$ and $V_1(\phi)$ have stationary points at $\phi = 0$ and $\phi = v$.  These points are respectively local maxima and minima of $V_0(\phi)$.  $V_1(\phi)$ has a saddle point at $\phi=v$ and either a maximum or a minimum at $\phi = 0$, depending upon the leading sign.

We obtain the free energy density  by adding the finite-temperature one-loop correction
\begin{equation}
\label{eq:F1}
\mathcal{F}_1(\phi, T) = \sum_i \frac{g_i T^4}{2\pi^2} I_\mp \left[\frac{m_i(\phi)}{T}\right]
+ \sum_\text{bosons} \frac{Tg}{12\pi}\left[m_i(\phi)^3 - [m_i(\phi)^2 + \Pi_i(T)]^{3/2}\right],
\end{equation}
where $I_-$ and $I_+$ are for the relevant thermal distribution functions for the bosonic and for the fermionic contributions, respectively:
\begin{equation}
\label{eq:I}
I_\mp (x) = \pm \int_0^\infty dy\; y^2 \log \left(1\mp \exp^{-\sqrt{y^2+x^2}}\right),
\end{equation}
and where the second summation is for the resummed Daisy diagrams with $\Pi_i(T) = \frac{1}{3}h_i^2 T^2$, where $h_i$ is the Yukawa coupling.
Assuming all particles acquire mass through a Higgs-like mechanism in which the mass terms are of the form $m_i(\phi) = h_i\phi$ (i.e. neglecting explicit mass terms for all the additional degrees of freedom interacting with the field $\phi$), the free energy density takes the form
\begin{multline}
\label{eq:F}
\mathcal{F}(\phi, T) =  \lambda(-\tfrac{1}{2} v^2 \phi^2 + \tfrac{1}{4} \phi^4)  
+ \sum \pm \frac{g_i h_i^4}{64\pi^2}\left[ \phi^4 \left(\log\frac{\phi^2}{v^2} - \frac{3}{2}\right) + 2v^2\phi^2\right] \\
%+ \sum \frac{g_i T^4}{2\pi^2} I_\mp \left(\frac{h_i \phi}{T}\right)
%+ \sum_\text{bosons} \frac{Tg_ih_i^3}{12\pi}\left[\phi^3 - (\phi^2 + \tfrac{1}{3}T^2)^{3/2}\right] \\
+ \sum \frac{g_i T^4}{2\pi^2} \left[ I_\mp \left(\frac{h_i \phi}{T}\right) +
\frac{\pi(-1\mp1)}{12} %\left\{ \left(\frac{h_i\phi}{T}\right)^3 - \left[ \left(\frac{h_i\phi}{T}\right)^2 + \frac{h^2}{3} \right]^{3/2} \right\}\right]
D\left(\frac{h_i\phi}{T}, h_i\right)\right]
 \end{multline}% (see~Ref. \cite{Megevand:2007sv}).
 where $D(x,h) = (x^2 + \tfrac{1}{3}h^2)^{3/2} - x^3$.
 For our purposes, we can ignore a constant vacuum energy term.  It will also be interesting in our analysis to add a temperature independent cubic term $V_0'(\phi) = \alpha (\tfrac{1}{2}v^2\phi^2 -\tfrac{1}{3}v\phi^3)$ with $\alpha \ll \lambda$ to account for example for the possible effects, at the level of the $\phi$ effective potential, of tree level cubic terms, driven by mixing with one or multiple gauge singlet scalar fields (see e.g. Ref.~\cite{Haber:1989xc}).
Notice that the purpose of the quadratic term in $V_0'$ is to ensure that the vacuum expectation value is maintained at $\phi = v$ at zero temperature.

%\subsubsection{Low and high-temperature expansions}

In the high-temperature (low-$\phi$) limit, Eq.~(\ref{eq:I}) can be expanded as
\begin{gather}
\begin{split}
\label{eq:I-high}
I_-(x) = &-\!\frac{\pi^4}{45} + \frac{\pi^2}{12}x^2 - \frac{\pi}{6}x^3 - \frac{x^4}{32}\log\frac{x^2}{a_b} \\
&-\!2\pi^{7/2} \sum_{l=1}^{\infty} (-1)^l \frac{\zeta(2l+1)}{(l+2)!}\Gamma\left(l+\tfrac{1}{2}\right)\left(\frac{x}{2\pi}\right)^{2l+4}, \text{ and}
\end{split} \\
\begin{split}
\label{eq:I+high}
I_+(x) = &-\!\frac{7\pi^4}{360} + \frac{\pi^2}{24}x^2  + \frac{x^4}{32}\log\frac{x^2}{a_f} \\
& +\!\tfrac{1}{4}\pi^{7/2} \sum_{l=1}^{\infty} (-1)^l \frac{\zeta(2l+1)}{(l+2)!}\left(1 - \frac{1}{2^{2l+1}}\right)\Gamma\left(l+\tfrac{1}{2}\right)\left(\frac{x}{\pi}\right)^{2l+4},
\end{split}
\end{gather}
(see e.g. the seminal work of Ref.~\cite{Anderson:1992}; see also \cite{Megevand:2007sv}) where $\log a_b = \tfrac{3}{2} - 2\gamma + 2\log(4\pi)$, $\log a_f = \tfrac{3}{2} - 2\gamma + 2\log\pi$, $\gamma$ is the Euler constant, $\zeta$ is the Riemann zeta function, and $\Gamma$ is the Gamma function.

In the low-temperature (high-$\phi$) limit, the expansions are given by
\begin{equation}
\label{eq:Ilow}
I_\mp (x) = -x^2  \sum_{k=1}^\infty \frac{(\pm1)^{k+1}}{k^2} K_2(kx),
\end{equation}
(again, see e.g. Ref.~\cite{Anderson:1992} and \cite{Megevand:2007sv}) where $K_2$ is the modified Bessel function of the second kind of order 2.

\begin{figure}[t]
   \centering
   \includegraphics{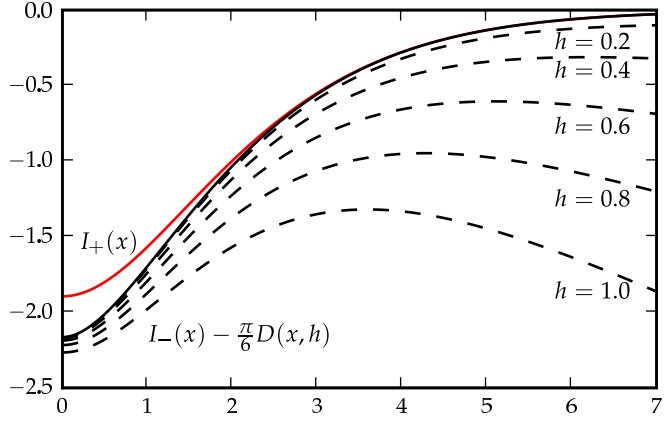}
   \caption{\it\small The integrals of Eq.~(\ref{eq:I}).  Bosons correspond to $I_-(x)$ (black line), and fermions correspond to $I_+(x)$ (red line). The dashed lines show the resummed Daisy diagram contributions
%, $I_-(x) - \frac{\pi}{6}D(x, h)$, 
for $h = $0.2, 0.4, 0.6, 0.8, and 1.0.}
   \label{fig:I+-}
\end{figure}

There are a few important things to note about these expansions.  The original integrals in Eq.~(\ref{eq:I}) are negative monotonically increasing functions with $\lim_{x\to\infty} I_\pm(x) = 0$.  They are similar in shape to upside-down bell-curves (see Fig.~\ref{fig:I+-}).  The low-temperature expansions have this same character for any finite number of terms.  If we keep only terms up to $k=8$ (as we do for the numerical study we present in this analysis), the errors at $x=0$ are only 0.01\% and 0.05\% for $I_+$ and $I_-$, respectively.  The errors drop exponentially for $x>0$, with fractional errors of $\sim 10^{-10}$ at $x = 2$.  In contrast, the high-temperature expansion diverges towards $\pm\infty$ for a finite number of terms.  For example, retaining even up to $l = 15$ produces a visible divergence in $I_+(x)$ towards negative infinity at $x=3.5$.  The benefit of the high-temperature expansions is that they contain an explicit cubic term, whereas the low-temperature expansions do not.  If the phase transition is weakly first-order so that terms of order $x^5$ can be ignored, then the cubic term is necessary to avoid a continuous phase transition.  Therefore, the low-temperature expansions underestimate the strength of the transition when the transition is weak.

As argued in Ref.~\cite{Espinosa:2008kw}, higher-order corrections to the potential are subdominant and can be safely neglected. In particular, Ref.~\cite{Espinosa:2008kw}, which adopts a setup very similar to ours, finds that two-loop effects, as caluclated e.g. in \cite{Espinosa:2000df}, do not affect significantly the structure of the effective potential. We thus neglect them here.

%%%%%%%%%%%%%%%%%%%%%%%%%%%%%%%

\subsection{Semi-Analytical Results in the High-Temperature Limit}
\label{sec:semi-analytic}

If the phase transition is weakly first-order, we may use a temperature-dependent quartic potential as an approximation to the free energy:
\begin{equation}
\label{eq:Fhigh}
\mathcal{F}(\phi, T) = D(T^2 - T_0^2)\phi^2 + (\tfrac{1}{3}\alpha v - ET)\phi^3 + \tfrac{1}{4}\lambda \phi^4 - \frac{\pi^2}{90}g_l T^4.
\end{equation}
The constants $D$, $E$, and $T_0$ come directly from the coefficients of Eq.~(\ref{eq:F}) and Eqs.~(\ref{eq:I-high}-\ref{eq:I+high}), and $g_l = g_b + \tfrac{7}{8}g_f$ are the effective number of degrees of freedom of particle species.  For this analysis we assume that $D$ is positive, which does not necessarily follow from Eq.~(\ref{eq:F}).
% $e=\alpha v/3$
%$x = \frac{\alpha v}{3ET_0}$

It is much more simple to analyze this equation if we recast it in the dimensionless form
\begin{equation}
\label{eq:Fhigh2}
\tilde{\mathcal{F}}(\varphi, \tau) = D(\tau^2 - 1) \varphi ^2 + E(x - \tau) \varphi ^3 + \tfrac{1}{4}\lambda \varphi^4 - \frac{\pi^2}{90}g_l \tau^4,
\end{equation}
where $\tilde{\mathcal{F}} = \mathcal{F}/T_0^4$, $\varphi = \phi/T_0$, $\tau = T/T_0$, and $x =\tfrac{1}{3}{\alpha v}/{ET_0}$.
%$x = e/ET_0$.
The critical temperature (the temperature of degenerate minima), is then easy to find analytically:
\begin{equation}
\label{eq:Tc}
\tau_c = \frac{- x \pm \sqrt{y\left[ y - \left(1 - x^2\right) \right]} }{y-1},
\end{equation}
where $y = D\lambda/E^2$.  In all situations, the solution corresponding to the positive root is the correct physical solution, while the negative root is an unphysical solution resulting from the use of an approximate potential.  Note that if $x > 1$ then by Eq.~(\ref{eq:Tc}) $x>\tau_c$ and the second minimum is at $\varphi < 0$ at $\tau = \tau_c$.  Since the temperature-dependent terms in Eq.~(\ref{eq:F}) are even in $\phi$ while those in Eq.~(\ref{eq:Fhigh}) are not, we reject all solutions with $\phi < 0$ and demand that $x < 1$.

If the transition temperature is close to the critical temperature, as we expect for a weakly first-order phase transition, then Eq.~(\ref{eq:dilution1}) gives the correct amount of dilution.  The entropy density is $s = -{d\mathcal{F}}/{dT} \rightarrow s/T_0^3 = -d\tilde{\mathcal{F}}/d\tau$.  In the high-temperature phase at the critical temperature, we have $s_+/T_0^3 = \frac{2\pi^2}{45}g_l\tau_c^3$.  In the low temperature phase, the minimum is 
$\varphi_c = 2E(\tau_c-x)/\lambda$ and the entropy difference is $\Delta s/T_0^3 = \varphi_c^2(2D\tau_c - E\varphi_c)$.  Plugging in values, we get
\begin{equation}
\frac{\Delta s}{s_+} = 
\frac{45}{2g_l\pi^2} \frac{8E^4}{\lambda^3} \left( y - \frac{y - x\sqrt{(x^2 + y-1)y}}{x^2+y} \right)
	\left(\frac{y - x\sqrt{(x^2 + y-1)y}}{x^2+y} \right)^2,
\end{equation}
which simplifies to 
\begin{equation}\label{eq:dilut}
\frac{\Delta s}{s_+} \approx
\frac{45}{2g_l\pi^2} \frac{8DE^2}{\lambda^2},
\end{equation}
in the limit that $y \gg x, 1$.  Note that the cubic term does not have a noticeable effect upon the dilution until it becomes quite large, which would violate our original assumptions.

In passing, we remark that, taking the above analysis at face value, in the Standard Model, where the top quark and the $W$ and $Z$ bosons are the only relevant degrees of freedom (see Ref.~\cite{Quiros:1999jp}), and $D = 0.16$, $E = 0.0096$, $y \sim 300$ (dependent upon the Higgs mass), and $g_l = 106.75$, one finds a negligible dilution factor, namely:
\begin{equation}
\left(\frac{a_f}{a_i}\right)^3 = \frac{s_+}{s_-} \approx 1 + \frac{\Delta s}{s_+}=1+\frac{45}{2g_l\pi^2} \frac{8DE^2}{\lambda^2}\approx 1.0001.
\end{equation}

Interestingly, the semi-analytic setup outlined above also allows us to get an estimate of the dilution factor expected in the context of the minimal supersymmetric extension of the Standard Model with light scalar tops (see e.g. Ref.~\cite{Carena:2002ss, Konstandin:2005cd, Carena:2008vj}), which has often been considered in the context of electro-weak baryogenesis. In that case, the cubic term $E$ can be one order of magnitude larger than in the Standard Model, implying a dilution factor of at most 1.01. According to our equation (\ref{eq:dilut}) this means that in the MSSM the dilution to the relic density of species freezing out prior to the electro-weak phase transition is a small effect (at most a few percent).

%%%%%%%%%%%%%%%%%%%%%%%%%%%%%%%

%\subsection{Models with Additional Scalar Degrees of Freedom}
\subsection{Strongly First-order Phase Transitions: Numerical Solutions}
\label{sec:numeric-dilution}

If the phase transition is strongly first-order, Eq.~(\ref{eq:Fhigh}) does not hold and we must instead resort to the full expression for the free energy density with the low-temperature expansion of Eq.~(\ref{eq:I}).  One must compute the dilution factor fully numerically.  

In our models, we proceed with the following steps.  First, we obtain a function of the minimum of the true vacuum $\phi_0(T)$ up to the critical temperature by numerically solving the differential equation
\begin{equation}
\frac{d\phi_0}{dT} = -\left.\left(\frac{\partial^2\mathcal{F}}{\partial \phi \partial T}\right) \right/ \left(\frac{\partial^2\mathcal{F}}{\partial\phi^2}\right)
\end{equation}
with the initial condition $\phi_0(T=0) = v$.  We then calculate the transition temperature $T_*$ by calculating the Euclidean action at many temperatures and searching for $S_3/T \sim \mathcal{O}$(130--140).  To account for a potentially wide range of transition temperatures and relativistic degrees of freedom, we use  $S_3(T_*)/T_* = 170 - 5\log(T/\text{1 GeV}) - 2\log(g)$.  We then calculate the dilution factor directly from Eq.~(\ref{eq:dilution3}) using $s = -d\mathcal{F}/dT$.

\begin{figure*}[t]
\includegraphics[width=3.0in]{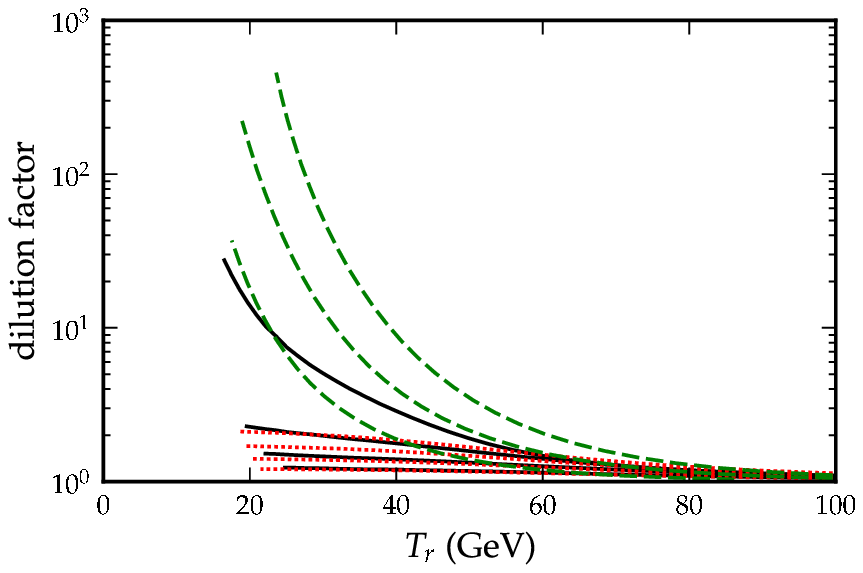}
\includegraphics[width=3.0in]{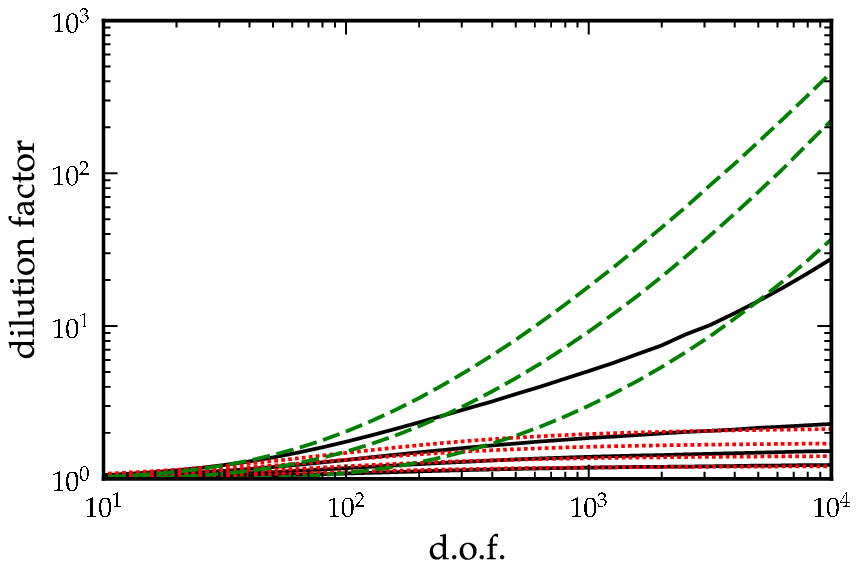}
\caption{\label{fig:dilution} \it Dilution factors for boson-fermion models (dotted red and dashed green) and purely bosonic models (solid black) as functions of the temperature at the end of reheating (left) and the number of fermion and boson degrees of freedom (right).  For the purely bosonic models, the highest line corresponds to the highest Yukawa coupling for which the point at $\phi=0$ is a minimum at zero temperature.  The three lower lines have 0.7, 0.8 and 0.9 times this value.  For boson-fermion models, the four dotted red lines correspond to Yukawa couplings identical to those in the bosonic cases, while the dashed green lines corresond to Yukawa couplings of 0.5, 0.75, and 1.0.}
\end{figure*}

\begin{figure*}[t]
\centering
\includegraphics [width=3.0in]{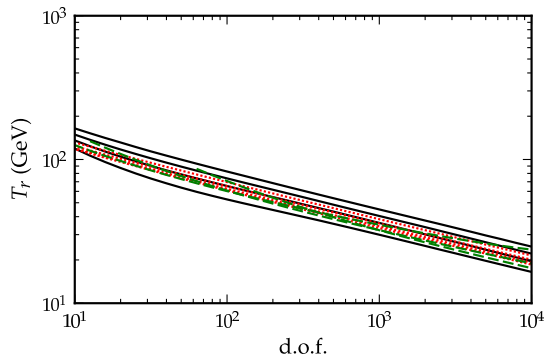}
\caption{\label{fig:Trvsg}\it\small Reheating temperature as a function of fermion and boson degrees of freedom.  Lines correspond to the same models as in Fig.~\ref{fig:dilution}, with higher Yukawa couplings resulting in lower reheating temperatures.}
\end{figure*}

For definiteness, we consider two classes of models: one with only additional scalar bosonic degrees of freedom, and one with an equal number of additional bosonic and fermionic contributions.  An explicit realization of the former case is given by the multiple hidden sector scalars $S_i$ discussed in Ref.~\cite{Espinosa:2007qk,Espinosa:2008kw}, where the coupling between the hidden and the visible sector occurs only through renormalizable terms in the potential proportional to $H^\dagger H S_i^2$. The latter, instead, alludes to a supersymmetric particle content for the additional degrees of freedom; one possible example is an extension to multiple extra generations of fermion-sfermions along the lines of the analysis of Ref.~\cite{Fok:2008yg}, that discusses the impact on the EWPT of a supersymmetric model with four chiral matter generations.

For the ease of analysis, all massive particles in a given model have the same Yukawa coupling so that we effectively have only one (or two) massive particle species with a potentially large number of degrees of freedom.  In each case, the degrees of freedom per fermion and/or boson range from 10 to 10,000.  We additionally add massless bosons (i.e., $h=0$) with $g_l = 100$ degrees of freedom, which approximates the number of relativistic degrees of freedom in the standard model.
That is, we have a variable number of degrees of freedom for massive particles, but keep the number of massless particles constant.  The latter just adds the $\phi$-independent term $-{\pi^2g_l}T^4/90$ to the free energy of Eq.~(\ref{eq:F}), and gives an additional suppression to the factor $\zeta$ of Eq.~(\ref{eq:T*}).  Note that the addition of massless particles does not change the phase transition temperature or dynamics.  It does, however, add entropy to both the high and low-temperature phases, which decreases the overall dilution factor via Eq.~\ref{eq:dilution3}.
% {\bf MAX:CHECK THIS} Both effects are mild, if one considers the opposite regime where {\em all} of the additional degrees of freedom are massless (for instance, it is clear that this amounts to only a small logarithmic correction in Eq.~(\ref{eq:T*})). 
%The value of $g_l$ is chosen to be close to that of the standard model.
In the boson-only models, large couplings turn the point at $\phi=0$ into a maximum at zero temperature via the quadratic term in Eq.~(\ref{eq:V1}).  We limit ourselves to the maximum couplings for which this is not the case.  In all cases, we set the zero-temperature Higgs mass to be $m_h = 150$ GeV.

Figures~\ref{fig:dilution} and~\ref{fig:Trvsg} display our results.  The black lines correspond to models without any fermion contribution and variable Yukawa couplings.  The models in the highest of these lines have the maximal couplings for which the point at $\phi = 0$, $T = 0$ is a maximum, while the lower lines have 0.9, 0.8, and 0.7 times the maximum couplings.  The dotted red and dashed green lines correspond to models with equal fermion and boson contributions.  The dotted red lines have the same couplings as the solid black lines, while the dashed green lines have fixed couplings of 0.5, 0.75, and 1.0.
For boson-fermion models with reasonably large Yukawa couplings and many degrees of freedom, one can achieve dilution factors on the order of 100.  Purely bosonic models, on the other hand, can only reach dilution factors on the order of 10.  
The largest aspect contributing to this discrepancy is the range of Yukawa couplings available to the different cases.  At $g=$ 10,000, the maximum coupling for bosons is only $h_b = 0.23$, while the highest plotted dilution factor for a boson-fermion model has $h_b = h_f = 1.0$.  When purely bosonic and boson-fermion models have identical couplings, the bosonic models tend to have slightly higher dilution factors.

One can easily explain two of the important qualitative features of Figs.~\ref{fig:dilution} and~\ref{fig:Trvsg} by examining the free energy of Eq.~(\ref{eq:F}).  First, the temperature of the phase transition tends to decrease with increasing degrees of freedom.  The temperature-dependent term in the free energy has a coefficient of $gT^4$.  When this term is large, the second minimum of the potential disappears and the origin becomes the true vacuum.  Thus, we expect the temperature of the transition to scale as $T \propto g^{-1/4}$, as seen in Fig.~\ref{fig:Trvsg}.
Second, the dilution factor tends to increase with increasing Yukawa couplings and decreasing temperature scales.  If either $h$ is large or $T$ is small, then the temperature-dependent term is nearly zero except for small $\phi$, excluding resummed Daisy terms (see Fig.~\ref{fig:I+-} for the behavior of $I_\pm(x)$).  Therefore, the entropy $s = -d\mathcal{F}/dT$ is much smaller in the true vacuum than it is at the origin.  Including the Daisy terms, the negative slope of $I_-(x) - \tfrac{\pi}{6}D(x,h)$ tends to decrease the entropy at high $\phi$.
If the transition reaches a phase coexistence stage, then the small entropy leads directly to a high dilution factor via Eq.~(\ref{eq:dilution1}).  This effect dominates when $h\phi/T \approx hv/T \gtrsim 5$, which is the case for the boson-fermion models in Fig.~\ref{fig:dilution} with $h \geq 0.5$ (dashed green lines).

%%%%%%%%%%%%%%%%%%%%%%%%%%%%%%%

\subsection{Models with Tree-Level Cubic Terms}

\begin{figure}[t]
\centering
\includegraphics[width=3.0in]{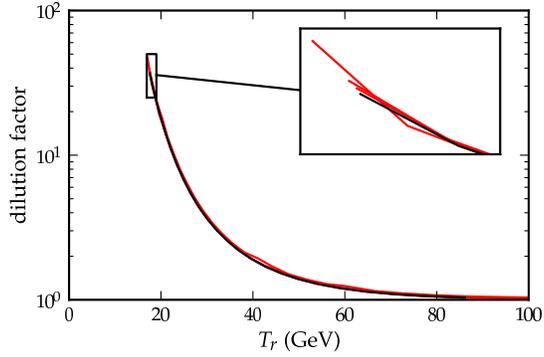}
\caption{\label{fig:cubic}\it\small Dilution factors for models with cubic terms (red lines) added to the boson-fermion models in Fig.~\ref{fig:dilution} with Yukawa couplings of $h = 0.5$ (black line).  The red lines have cubic strengths of $\alpha/\lambda =$ 0.05, 0.15, and 0.50.}
\end{figure}

We also examine what happens when we add a tree-level cubic term $V_0'(\phi) = \alpha (\tfrac{1}{2}v^2\phi^2 -\tfrac{1}{3}v\phi^3)$ to the free energy.  Fig.~\ref{fig:cubic} shows cubic terms added to boson-fermion models with Yukawa couplings fixed at $h_f = h_b =0.5$.  The models are otherwise exactly the same as those in the previous section: $g_l = 100$, $m_h = 150$ GeV, and the boson/fermion degrees of freedom $g_f = g_b$ range from 10 (corresponding to the lower-right portion of Fig.~\ref{fig:cubic}) to 10,000.
The black line has no added cubic term, while the red lines have cubic strengths of  $\alpha/\lambda =$ 0.05, 0.15, and 0.50.
 Even large terms with $\alpha = \tfrac{1}{2}\lambda$ do not seem to significantly impact the dilution factor.  The effects of cubic terms upon purely bosonic models and models with smaller Yukawa couplings are qualitatively similar. 

%%%%%%%%%%%%%%%%%%%%%%%%%%%%%%%%%%%%%%%%%%%%%%%%%%%%%%%%%%%%%%%%%%%%

\section{Supersymmetric Dark Matter and a Strongly First-Order Phase Transition}
\label{sec:EWPT-SUSY}

\begin{figure}[t]
\centering
\includegraphics[width=3.0in]{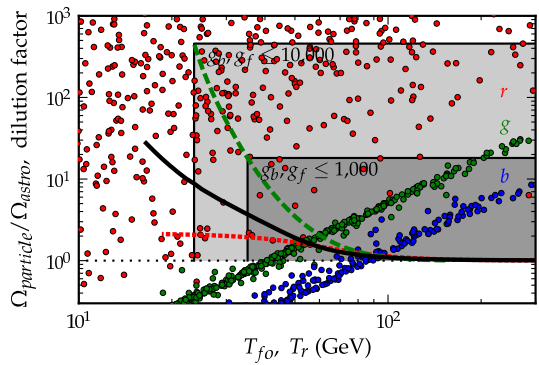}
\caption{\label{fig:combo}\it\small An overlay of dilution factors for boson and boson-fermion models on top of a scatter plot of dark matter abundance in different MSSM models.  The points are the same as those in Fig.~\ref{fig:mssm}.  The green, black, and red lines correspond to the highest green, black, and red lines of Fig.~\ref{fig:dilution}.  The two boxes show which MSSM models are viable for dilution models with $h_f=h_b=1.0$ and $g_b, g_f \leq$ 10,000 and 1,000.}
\end{figure}

We determine here which MSSM models with neutralino dark matter might not overproduce dark matter given a strongly first-order phase transition.  Figure~\ref{fig:combo} shows an overlay of the dilution from different models of phase transitions on top of our scan of the dark matter abundance in MSSM models.  Red, green and blue points correspond to bino, higgsino, and wino-like models, while the solid black, dotted red, and dashed green lines correspond to phase transition models with only bosons at their maximal Yukawa couplings, bosons and fermions at the same Yukawa couplings, and bosons and fermions with fixed Yukawa couplings of $h_f = h_b = 1.0$.  These are the same as the highest of the black, red and green lines in Fig.~\ref{fig:dilution}.

The freeze-out temperature of the dark matter must be larger than the final reheating temperature of the phase transition if the transition is to have an effect upon dark matter abundance.  Therefore, any point that lies to the right of any of the three lines can be diluted to the cosmic relic abundance by a sufficiently strong phase transition.  If a point lies between the lines, then one can easily pick a phase transition model that produces the correct amount of dilution.  Note that only MSSM models with an overabundance of dark matter are problematic---if a given phase transition model over-dilutes a given MSSM model of dark matter, then the two models can still be mutually viable if the neutralino dark matter does not constitute the entire universal dark matter abundance.

The two boxes in Fig.~\ref{fig:combo} show which MSSM models are viable given the most optimistic phase transition models (that is, boson-fermion models with $h_f=h_b=1.0$) with degrees of freedom $g_f = g_b =$ 1,000 (inner box) and 10,000 (outer box).  With 1,000 d.o.f., almost all of the over-abundant higgsino and wino-like models are viable.  With 10,000 d.o.f., a handful of bino-like models become viable as well.  However, there are no viable models with freeze-out temperatures less than $\sim 20$ GeV or overabundances greater than a factor of $\sim 200$  unless they are accompanied by phase transitions with very large ($>$ 10,000) particle degrees of freedom.  This includes most of the bino-like models and a small subset of the low-mass higgsino-like models.

\section{Summary and Conclusions}\label{sec:concl}
A strongly first-order electro-weak phase transition, warranted in the context of scenarios for the production of the observed baryon asymmetry at the electro-weak scale, can lead to a significant dilution of the thermal relic abundance of the lightest supersymmetric particle, if the latter is stable. In this paper we studied the impact of toy models for the electro-weak sector on the relic abundance of MSSM neutralinos. Specifically, we considered the effect on the electro-weak sector of adding a large number of bosonic or bosonic-plus-fermionic degrees of freedom, which affects both the one-loop temperature-independent and the temperature-dependent contribution to the effective potential of the Higgs field. We noted that for neutralino freeze-out temperatures between 20-40 GeV, corresponding to neutralino masses at or above 400 GeV, the dilution due to the entropy injected in the first-order electro-weak phase transition can be as large as 10--100 if we postulate a large number of extra degrees of freedom ($10^3--10^4$). Numerous MSSM models exist that could be viable if such a dilution effect is in fact operative. For models with a more modest number of additional d.o.f. (say, of the same order of the Standard Model d.o.f.), the dilution can still be on the order of 2. We find that a cubic term in the effective potential has a comparatively small effect upon the relic abundance of species. Also, we showed that the dilution expected in the context of the MSSM without additional degrees of freedom coupled to the Higgs sector is a very small effect, at most at the few-percent level.

Should future data from the LHC point towards a neutralino relic abundance larger than the cosmological dark matter density, the origin of the mis-match might lie in the same electro-weak scale physics that the LHC itself will concurrently explore. Interestingly, this can be profoundly intertwined with the question of the origin of the matter-antimatter asymmetry in the early universe, opening up the possibility that electro-weak physics lie at the core of both dark and visible matter, and that the LHC will soon shed light and perhaps unveil this scenario.

\begin{acknowledgments}
We gratefully acknowledge useful conversation with Michael Dine and Anthony Aguirre. The suggestion to investigate this topic was originally proposed to SP by Hitoshi Murayama, to whom we are gratefully indebted. 

CW receives support from the Graduate Assistance in Areas of National Need program. SP is partly supported by an Outstanding Junior Investigator Award from the US Department of Energy under Contract DE-FG02-04ER41268, and by NSF Grant PHY-0757911.
\end{acknowledgments}

\end{document}